\pdfoutput=1
\documentclass[reprint,superscriptaddress,aip,apl,floatfix]{revtex4-2}

\usepackage[separate-uncertainty=true]{siunitx}
\usepackage{graphicx}

\usepackage[english]{babel}

\begin{document}

\title{A fast and large bandwidth superconducting variable coupler}

\date{November 18, 2020}

\author{H.-S. Chang}
\affiliation{Pritzker School of Molecular Engineering, University of Chicago, Chicago IL 60637, USA}
\author{K. J. Satzinger}
\altaffiliation[Present address: ]{Google, Santa Barbara CA 93117, USA.}
\affiliation{Department of Physics, University of California, Santa Barbara CA 93106, USA}
\affiliation{Pritzker School of Molecular Engineering, University of Chicago, Chicago IL 60637, USA}
\author{Y. P. Zhong}
\affiliation{Pritzker School of Molecular Engineering, University of Chicago, Chicago IL 60637, USA}
\author{A. Bienfait}
\altaffiliation[Present address: ]{Universit\'e de Lyon, ENS de Lyon, Universit\'e Claude Bernard, CNRS, Laboratoire de Physique, F-69342 Lyon, France}
\affiliation{Pritzker School of Molecular Engineering, University of Chicago, Chicago IL 60637, USA}
\author{M.-H. Chou}
\affiliation{Pritzker School of Molecular Engineering, University of Chicago, Chicago IL 60637, USA}
\affiliation{Department of Physics, University of Chicago, Chicago IL 60637, USA}
\author{C. R. Conner}
\affiliation{Pritzker School of Molecular Engineering, University of Chicago, Chicago IL 60637, USA}
\author{\'E. Dumur}
\altaffiliation[Present address: ]{Universit\'e Grenoble Alpes, CEA, INAC-Pheliqs, QuantECA, 38000 Grenoble, France}
\affiliation{Pritzker School of Molecular Engineering, University of Chicago, Chicago IL 60637, USA}
\affiliation{Argonne National Laboratory, Argonne IL 60439, USA}
\author{J. Grebel}
\affiliation{Pritzker School of Molecular Engineering, University of Chicago, Chicago IL 60637, USA}
\author{G. A. Peairs}
\affiliation{Department of Physics, University of California, Santa Barbara CA 93106, USA}
\affiliation{Pritzker School of Molecular Engineering, University of Chicago, Chicago IL 60637, USA}
\author{R. G. Povey}
\affiliation{Pritzker School of Molecular Engineering, University of Chicago, Chicago IL 60637, USA}
\affiliation{Department of Physics, University of Chicago, Chicago IL 60637, USA}
\author{A. N. Cleland}
\affiliation{Pritzker School of Molecular Engineering, University of Chicago, Chicago IL 60637, USA}
\affiliation{Argonne National Laboratory, Argonne IL 60439, USA}

\begin{abstract}
Variable microwave-frequency couplers are highly useful components in classical communication systems, and likely will play an important role in quantum communication applications. Conventional semiconductor-based microwave couplers have been used with superconducting quantum circuits, enabling for example the \emph{in situ} measurements of multiple devices via a common readout chain. However, the semiconducting elements are lossy, and furthermore dissipate energy when switched, making them unsuitable for cryogenic applications requiring rapid, repeated switching. Superconducting Josephson junction-based couplers can be designed for dissipation-free operation with fast switching and are easily integrated with superconducting quantum circuits. These enable on-chip, quantum-coherent routing of microwave photons, providing an appealing alternative to semiconductor switches. Here, we present and characterize a chip-based broadband microwave variable coupler, tunable over 4-8 GHz with over 1.5 GHz instantaneous bandwidth, based on the superconducting quantum interference device (SQUID) with two parallel Josephson junctions. The coupler is dissipation-free, features large on-off ratios in excess of 40 dB, and the coupling can be changed in about 10 ns. The simple design presented here can be readily integrated with superconducting qubit circuits, and can be easily generalized to realize a four- or more port device.
\end{abstract}

\maketitle

In recent years, superconducting qubit circuits have made significant progress, with the notable construction of a 53-qubit quantum processor \cite{arute2019}, as well as the exploration of high-fidelity quantum communication protocols using superconducting qubit circuits \cite{Kurpiers2018,Campagne2018,Axline2018,Leung2019,Zhong2019,Chang2020,Burkhart2020}, showing that distributed quantum computing using superconductors may be possible. This drives a need for active microwave components capable of routing and modulating single microwave photons while preserving quantum coherence, which could further enable the distribution of entanglement between different nodes in a quantum network \cite{Lang2013,Narla2016}. Additionally, such a coupler would allow for multiplexing of classical microwave signals, enabling the integrated readout and control of multiple qubits using a single microwave cable.

Conventional electromechanical microwave switches feature long switching times and dissipate significant amounts of heat when switched, features that are not ideal for cryogenic operation \cite{Ranzani2013}. Semiconductor-based microwave switches have been used in cryogenic environments \cite{Hornibrook2015}, but these also tend to be lossy, dissipate heat when switched, and are difficult to integrate with superconducting circuits. Superconducting switches, in contrast, allow for easy integration with superconducting qubits, afford dissipation-free operation, and can be switched on nanosecond time scales. A few different realizations of superconducting microwave-frequency couplers have been demonstrated \cite{Pechal2016,Naaman2016,Chapman2016,Schuermans2011,Abdo2017,Rosenthal2020}; however, these mostly have fixed operating frequencies, involve complicated fabrication, and often suffer from limited bandwidth.

Here we report the design, implementation and characterization of a broadband, three-port superconducting variable coupler, based on the Josephson flux-tunable DC SQUID \cite{clarke1977}.  We use the intrinsic $LC$ plasma resonance of the SQUID to control the transmission to each port in the coupler, enabling fast and widely-tunable operation, providing a simple approach that preserves a direct galvanic coupling through each port. We experimentally characterize the device at cryogenic temperatures, and demonstrate a large on-off power ratio of over 40 dB, a wide instantaneous bandwidth exceeding 1 GHz for both transmission and isolation, and fast switching times of about 10 ns. The design affords complete and continuous electronic control of the transmission between the three ports, enabling fast and coherent routing of single photon microwave signals. The coupler can be readily integrated with superconducting qubits, as shown in Ref.~\onlinecite{Chang2020}, and can be easily extended to a larger number of ports.

\begin{figure}
	\begin{center}
		\includegraphics[width=3.37in]{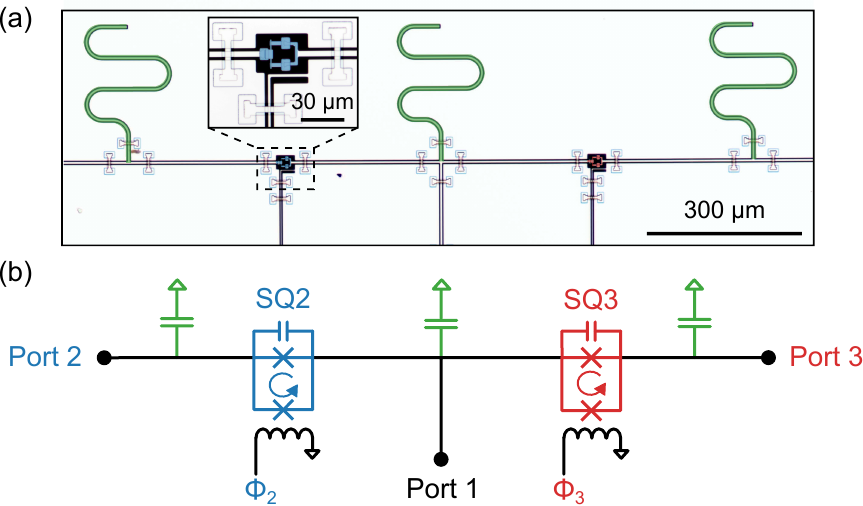}
	\end{center}
	\caption{
		\label{fig:fig1}
		Device description, where signals into port 2 or port 3 are routed to port 1, using those ports' associated SQUID couplers.
		(a) False-color optical micrograph of the complete device. Inset: magnified view of one SQUID and its associated flux control line.
		(b) Schematic circuit diagram for (a), with port 1 in black, port 2 and its associated SQUID (SQ2) in blue, port 3 and its SQUID (SQ3) in red, with three shunt capacitors in green.
	}
\end{figure}

Our three-port device controllably routes signals between ports 2 or 3 and port 1, using two Josephson SQUIDs as tunable resonators placed in series with ports 2 and 3, as shown in Fig.~\ref{fig:fig1}. Here measurements are all made with signals into port 2 and 3 and out of port 1; measurements with reversed signal direction is possible but was not measured here.  The SQUID plasma resonance frequency is defined as $\omega_p = 1/\sqrt{L C}$, where $L$ is the Josephson inductance of the SQUID, and $C$ is the associated parallel capacitance. When this frequency is resonant with the incoming signal, the SQUID presents a large series impedance, and fully reflects any incoming signals. Conversely, when the resonance frequency is far detuned from the incoming signal, it presents a small impedance and a substantial fraction of the incoming signal is transmitted through the SQUID. By controlling the magnetic flux $\Phi_{2}$ and $\Phi_{3}$ threaded through each SQUID using an on-chip flux line, we can tune the resonance frequency and thus vary the SQUID transmission continuously from zero to nearly unity. Using a shunting capacitance in parallel with the SQUID further improves the transmission in detuned operation, by better matching the circuit to the system impedance (here $Z_0 = 50~\Omega$).

The device is fabricated on a sapphire substrate with Al base wiring and Al-Al oxide-Al Josephson tunnel junctions, using a process outlined in Ref.~\onlinecite{Satzinger2018}. We characterize the coupler by cooling it in a dilution refrigerator with a base temperature below 10 mK and measuring the transmission between ports 2 and 3 and port 1 using a microwave vector network analyzer.  Details on the experimental setup, including the full wiring diagram, are provided in the Supplementary Information (SI \cite{SI}).

\begin{figure}[h!]
	\begin{center}
		\includegraphics[width=3.37in]{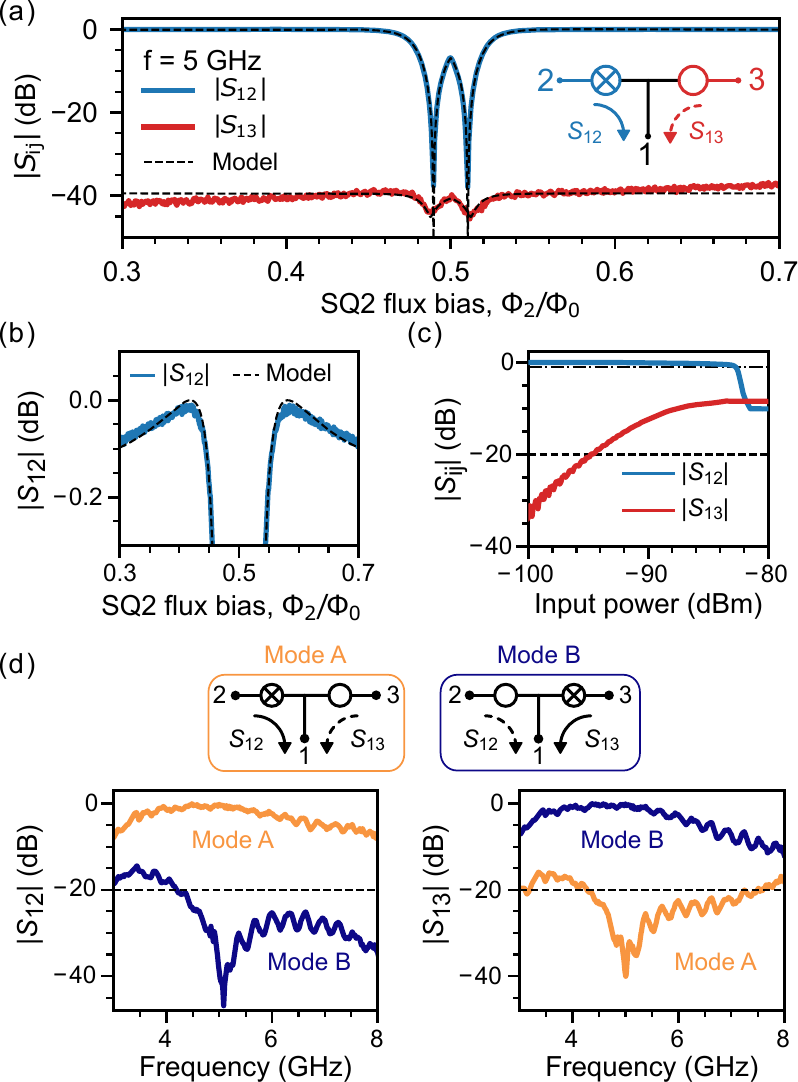}
	\end{center}
	\caption{
		\label{fig:fig2}
		Flux-controlled variable coupler transmission.
		(a) Normalized transmission for a fixed frequency signal at $f=5$ GHz from ports 2 and 3 to port 1, as a function of the SQ2 flux bias $\Phi_2$ in units of the magnetic flux quantum $\Phi_0 = h/2e$. Transmission $|S_{12}|$ is tuned continuously from unit transmission to -40 dB. Port 3's transmission $|S_{13}|$ is maintained near -40 dB for all SQ2 flux bias values. Dashed lines are a fit to a linear circuit model \cite{Pozar2011}, where the SQUIDs are modeled as parallel linear $LC$ resonators with flux-tunable inductances $L$.
		(b) Detail of the transmission vs. flux bias curve in (a), showing that near-unit transmission is achieved at two flux bias points, corresponding to where port 2's impedance is nearly matched to $50~\Omega$; see SI \cite{SI}.
		(c) Transmission $|S_{12}|$ (blue) and $|S_{13}|$ (red) at their maximum and minimum transmission points, respectively, as a function of port 2 and port 3 input power. Dashed and dash-dotted lines mark where transmission equals -20 dB and -1 dB, respectively.
		(d) Transmission $|S_{12}|$ (left) and $|S_{13}|$ (right) for two flux-tuning points, mode A (orange, see top left diagram) where port 2 is at maximum and port 3 at minimum transmission, and mode B (blue, see top right diagram) where these are reversed.  Dashed line represents the -20 dB transmission threshold used to define the isolation bandwidth.
	}
\end{figure}

In Fig.~\ref{fig:fig2}a, we show electronic control of transmission as a function of the control flux. With port 3 turned off ($|S_{13}| < -40$ dB), flux-biasing SQ2 controls the transmission $S_{12}$ from port 2 to port 1. The transmission can be tuned continuously from near-unit transmission ($\sim 0$ dB) to $-40$ dB, demonstrating a large on-off ratio (transmission calibration is described in the SI \cite{SI}). Transmission $S_{13}$ from port 3 to port 1 is maintained at a uniform small value near $-40$ dB for all SQ2 flux biases. We observe two pronounced dips in $|S_{12}|$ as a function of flux, which corresponds to where the SQUID plasma resonance is resonant with the probe signal at $f = 5$ GHz, reflecting almost all the signal. In Fig.~\ref{fig:fig2}b, we show a detail for the transmission at the two flux bias points where port 2's input impedance is nearly $50~\Omega$, resulting in near-unit transmission (the measured transmission is normalized to 0 dB at the maximum transmission point; see SI \cite{SI}). The coupler's insertion loss is characterized in a separate measurement, where it was found to be around $0.6 \pm 0.2$ dB (see SI \cite{SI}). We fit the flux dependence of the transmission using a linear circuit model \cite{Pozar2011}, modeling the SQUID as a parallel LC resonator with flux-tunable inductance; we extract the unbiased inductance of the SQUID to be $166.5 \pm 0.1$ pH, the SQUID capacitance to be $199.3 \pm 0.3$ fF, and the shunting capacitance to be $160.0 \pm 0.4$ fF.

At high powers, non-idealities due to the intrinsic SQUID nonlinearity become apparent. We characterize the power-handling of our device by measuring the transmission as a function of input power with port 2 ``on'' and port 3 ``off,'' as shown in Fig.~\ref{fig:fig2}c. To maintain a relative transmission level above $-1$ dB through port 2, we find a chip-level maximum power into port 2 of about -83 dBm, while maintaining the transmission below $-20$ dB for port 3, the maximum power into port 3 is about $-95$ dBm.

In Fig.~\ref{fig:fig2}d, we display the transmission through each port ($|S_{12}|$ and $|S_{13}|$) over a broad frequency range with each port tuned to its maximum and minimum transmission points respectively. In mode A, the SQUID fluxes are set so that port 2 is set to its maximum transmission while port 3 is set to its minimum, while in mode B the fluxes are set to where port 3's transmission is maximized while port 2's transmission is minimized. In mode A, we observe a flat transmission through port 2 with $|S^{A}_{12}| \gtrsim -1$ dB while port 3's transmission has $|S^{A}_{13}| \lesssim -20$ dB, both across a $\sim 1.5$ GHz band centered at 5 GHz. In mode B, we measure a large isolation bandwidth for port 2 with an equivalently broad and large transmission through port 3. The on-off ratios for both ports is greater than 40 dB.

These measurements demonstrate operation for a probe signal near 5 GHz; in the SI \cite{SI}, we show that the flux tunability of the SQUIDs allow equivalent operation for operating probe frequencies between 4 and 8 GHz, demonstrating tunable, broadband operation, achieving on-off ratios in excess of 40 dB, and ``on'' and ``off'' bandwidths of more than 1 GHz for all operating frequencies (see Table S1 in the SI \cite{SI}).

\begin{figure}
	\begin{center}
		\includegraphics[width=3.37in]{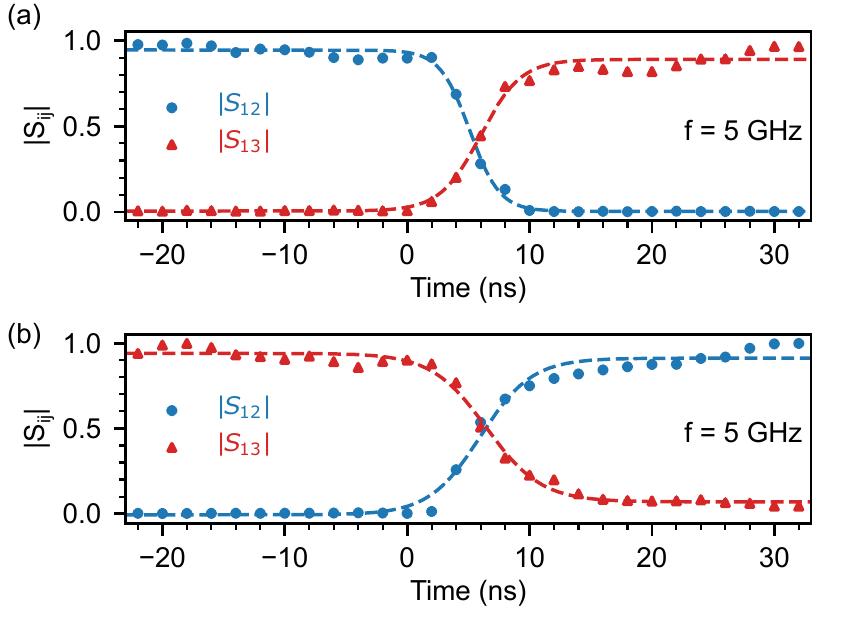}
	\end{center}
	\caption{
		\label{fig:fig3}
		Fast switching. Measured rising and falling edges of the waveforms through each port when switched between different operating states.
		(a) Mode A to mode B.
		(b) Mode B to mode A.
		Dashed lines are fit to a hyperbolic tangent to extract the rise and fall switching times.
	}
\end{figure}

We characterize the switching speed of the variable coupler by measuring the transient response of our device as it switches between the two modes, shown in Fig.~\ref{fig:fig3}. For these measurements, a fixed-frequency signal at $f=5$ GHz is sent to the coupler and a rectangular pulse is applied to each SQUID's flux line so as to switch from mode A to mode B or vice-versa.  The time-domain transmissions $|S_{12}|$ and $|S_{13}|$ are captured for each port using a fast analog-to-digital converter. By fitting the response waveform to a hyperbolic tangent, as shown in the figure, we extract $10\%$ to $90\%$ switching times of $8 \pm 1$ ns and $4.9 \pm 0.4$ ns for port 2's off-to-on and on-to-off times, respectively, and $6.6 \pm 0.8$ ns and $9.0 \pm 0.8$ ns for port 3's off-to-on and on-to-off times, respectively. These times are mostly limited by the bandwidth of our flux-bias electronics, which include a Gaussian filter with 250 MHz bandwidth.

\begin{figure*}[ht]
	\begin{center}
		\includegraphics[width=6.51in]{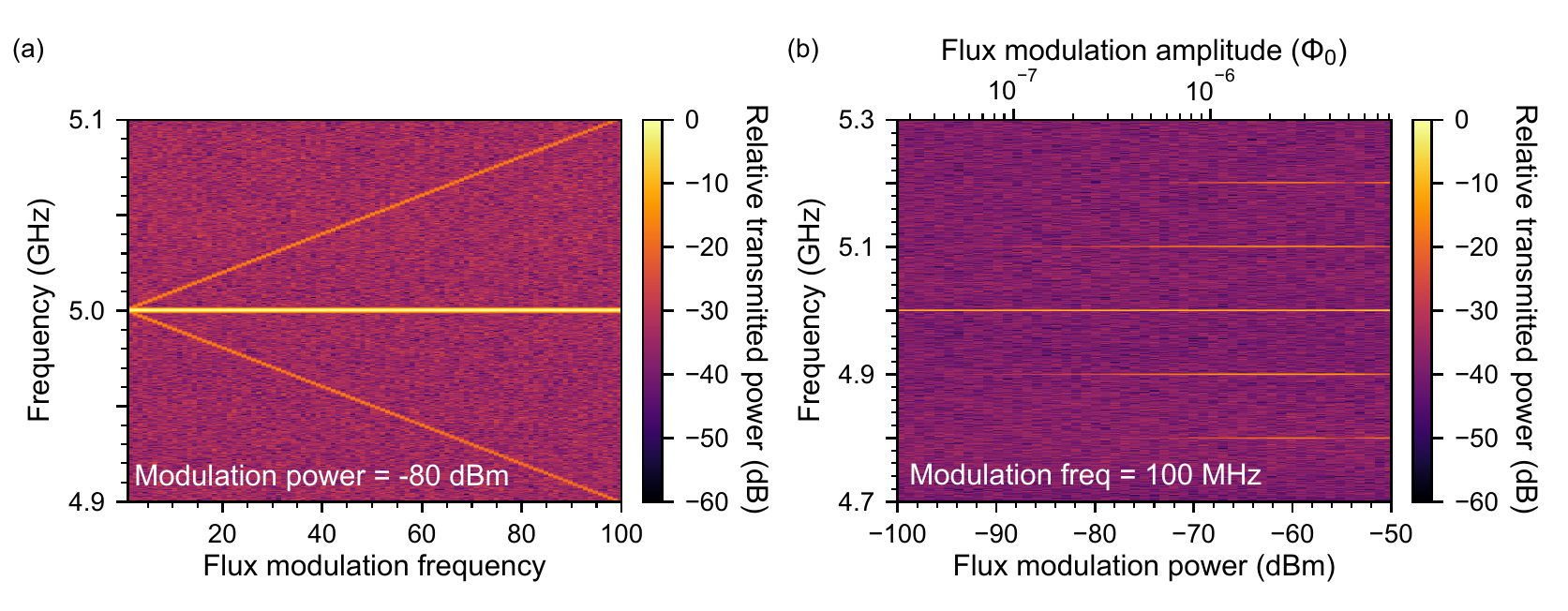}
	\end{center}
	\caption{
		\label{fig:fig4}
		Sideband generation.
		(a) Measured sideband spectrum as a function of the flux modulation frequency with a constant flux modulation power of -80 dBm. The measured spectrum is normalized by the transmitted power of the carrier signal at $f=5$ GHz.
		(b) Sideband spectrum as a function of flux modulation power, with equivalent modulation amplitude in units of the magnetic flux quantum $\Phi_0 = h/2e$ shown on top. The flux modulation frequency is kept constant at 100 MHz.
	}
\end{figure*}

A time-periodic modulation of the SQUID flux bias generates sidebands in the transmitted signal, offset from the carrier frequency by integer multiples of the modulation frequency. SQ2's flux bias is set to 0.473$\Phi_{0}$ where $|S_{12}|=-2.0$ dB and 0.052$\Phi_{0}$ from the maximum transmission bias, and port 3 is set to minimum transmission. A fixed-frequency signal at $f=5$ GHz is applied to port 2 as the carrier signal and SQ2's flux bias is modulated about its set-point with a small radiofrequency modulation of -80 dBm, corresponding to a modulation amplitude of $(2.6 \times 10^{-7}) \Phi_{0}$. We use a spectrum analyzer to measure the transmitted spectrum at port 1, which reveals sidebands on the carrier signal as shown in Fig.~\ref{fig:fig4}a as a function of the modulation frequency, as the latter is varied from 10 to 100 MHz. In Fig.~\ref{fig:fig4}b, we explore the dependence of these sidebands on the modulation power for a fixed modulation frequency, where at low powers only the carrier signal is detected, and as the power is increased the first and then additional sidebands appear in the spectrum. This sideband generation capability of the coupler could be useful for shifting the operating frequency or for frequency-domain multiplexing of microwave signals, for example for qubit control and readout.

In conclusion, we have presented and demonstrated a simple three-port superconducting variable coupler, using flux-biased SQUIDs operating at microwave frequencies to control the port-to-port transmission. The variable coupler has bandwidths in excess of 1 GHz, greater than 40 dB on-off coupling, and affords fast switching of a few nanoseconds. The flux tunability allows the center frequency to be set in the range of 4 to 8 GHz. The design can be easily modified to more than three ports, and can be readily integrated with superconducting qubit circuits for the routing and modulation of individual microwave photons \cite{Chang2020}.

\section*{Data Availability Statement}
The data that support the findings of this study are available from the corresponding author upon reasonable request.

\section*{Acknowledgements}
The authors thank P. J. Duda for helpful discussions. Devices and experiments were supported by the Air Force Office of Scientific Research and the Army Research Laboratory. K.J.S. was supported by NSF GRFP (NSF DGE-1144085), \'E.D. was supported by LDRD funds from Argonne National Laboratory; A.N.C. was supported in part by the DOE, Office of Basic Energy Sciences. This work was partially supported by the UChicago MRSEC (NSF award DMR-201185) and made use of the Pritzker Nanofabrication Facility, which receives support from SHyNE, a node of the National Science Foundation's National Nanotechnology Coordinated Infrastructure (NSF ECCS-2025633).

\bibliography{bibliography}

\end{document}


\begin{center}
	\textbf{\Large{Supplementary Information for A fast and large bandwidth superconducting variable coupler}}
\end{center}

\setcounter{figure}{0}   
\renewcommand{\thefigure}{S\arabic{figure}}
\renewcommand{\thetable}{S\arabic{table}}
\renewcommand{\theequation}{S\arabic{equation}}

\section{Device and experimental setup}
\begin{figure}
	\begin{center}		
	   \includegraphics[width=0.95\textwidth]{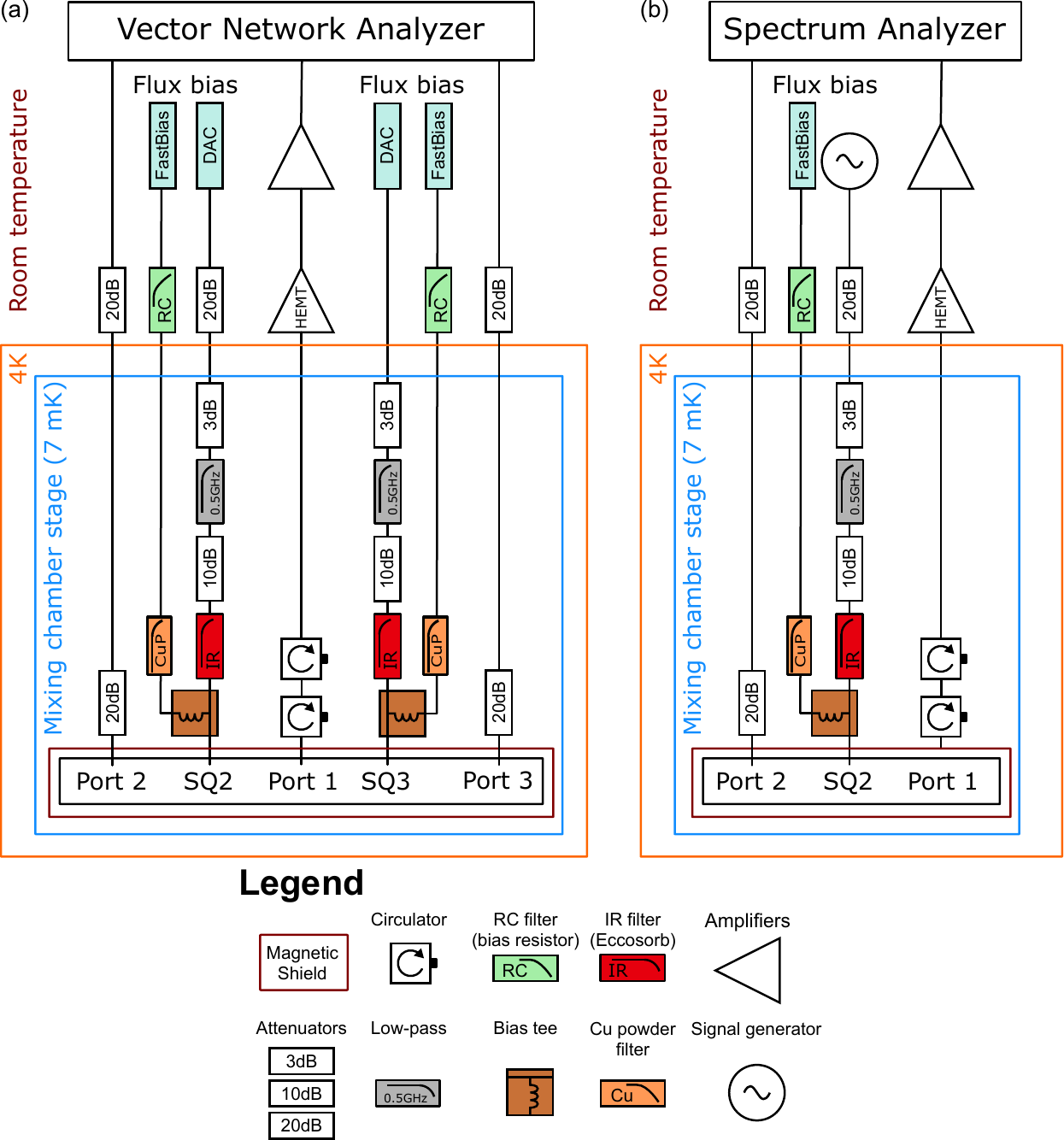}
    \end{center}
	\caption{
		\label{fig:wiring}
		Schematic of the measurement setups for measuring (a) microwave transmission through the device with a vector network analyzer and (b) sideband spectrum as shown in Fig.~4 of the main text with a spectrum analyzer. For this measurement, we connect the flux bias lines to a signal generator outside of the fridge to modulate the flux through the SQUID loop (SQ2). The output spectrum is measured with a spectrum analyzer. Solid lines mark the different temperature stages inside the dilution refrigerator.
	}
\end{figure}

The experiment is carried out inside a dilution refrigerator with a base temperature below 10 mK. Fig.~\ref{fig:wiring} shows the schematic of the wiring inside the refrigerator. Ports 2 and 3 are treated as input lines and are heavily attenuated and filtered at each temperature stage. Port 1 is treated as an output line, connected to two circulators in series (QuinStar) at the mixing temperature stage, then amplified by a high electron mobility transistor amplifier (Low Noise Factory) at a temperature of 4 K, followed by a room-temperature amplifier (Miteq). This wiring configuration allows us to measure the transmission from port 2 to port 1 and from port 3 to port 1 ($S_{12}$ and $S_{13}$) but not the transmission in the reverse direction ($S_{21}$ and $S_{31}$). The device is placed inside an aluminum sample box which is placed inside a single layer of high permeability magnetic shielding, bolted to the mixing chamber stage; electrical connections are made with aluminum wire bonds.

\clearpage
\section{Insertion loss}
\begin{figure}[h!]
	\begin{center}
		\includegraphics[width=0.8\textwidth]{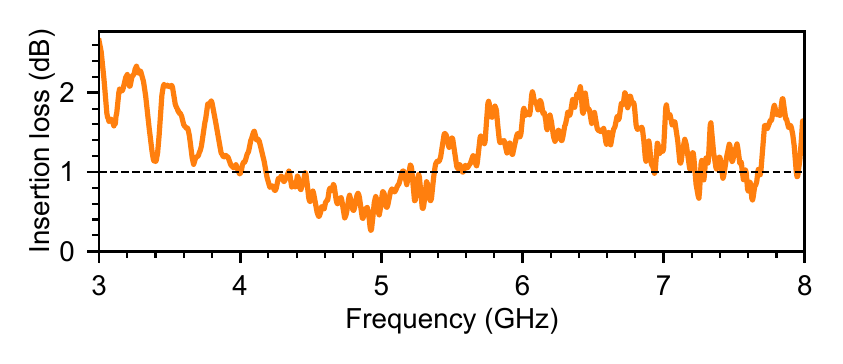}
	\end{center}
	\caption{
		\label{fig:insertion}
		Insertion loss of the device. Dashed horizontal line represents the 1 dB insertion loss point used to extract a bandwidth of 800 MHz.
	}
\end{figure}

The insertion loss of the variable coupler is measured by normalizing the measured transmission $S_{12}^a$ from port 2 to port 1 with port 2 set to maximum transmission and port 3 set to minimum transmission. This measurement was then compared with a reference measurement $S_{12}^b$ made in a subsequent cool-down by removing the device and connecting the port 2 and port 1 microwave lines directly to one another with a female/female SMA adapter. The insertion loss $L = -20\log_{10} \left (|S_{12}^a/S_{12}^b| \right )$ is then calculated as the ratio of these two transmission measurements, shown in Fig.~\ref{fig:insertion}. We observe a low insertion loss of less than 1 dB over a bandwidth of 800 MHz.

\section{Frequency tunability}
To demonstrate the broad tunability of the device, we measure the variable coupler at different operating frequencies ranging from 4 to 8 GHz and perform spectroscopy measurements similar to those in Fig.~2d of the main text. Device parameters extracted from these spectra are summarized in Table~\ref{table:performance}, and are similar to those in the main text.

\begin{figure}
	\begin{center}
		\includegraphics[width=0.95\textwidth]{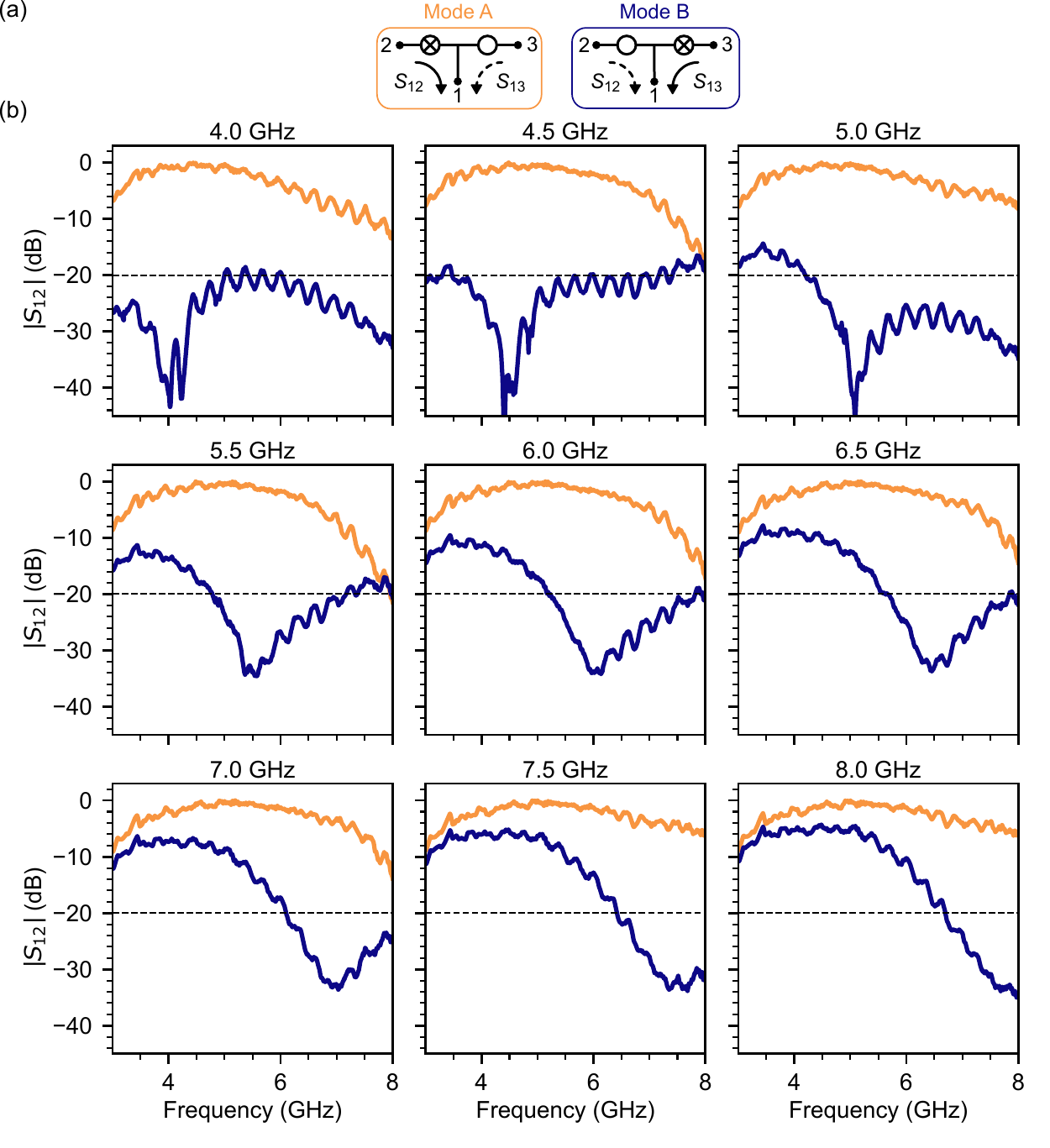}
	\end{center}
	\caption{
		\label{fig:s12all}
		Coupler performance for different operating frequencies, between ports 1 and 2.
		(a) Schematic of the two modes measured here; see main text.
		(b) Measured transmission from port 2 to port 1 $(|S_{12}|)$ as a function of frequency for mode A (orange) and mode B (blue), demonstrating broadband operation. Dashed line represents the $-20$ dB transmission threshold used to define the isolation bandwidth.
	}
\end{figure}

\begin{figure}
	\begin{center}
		\includegraphics[width=0.95\textwidth]{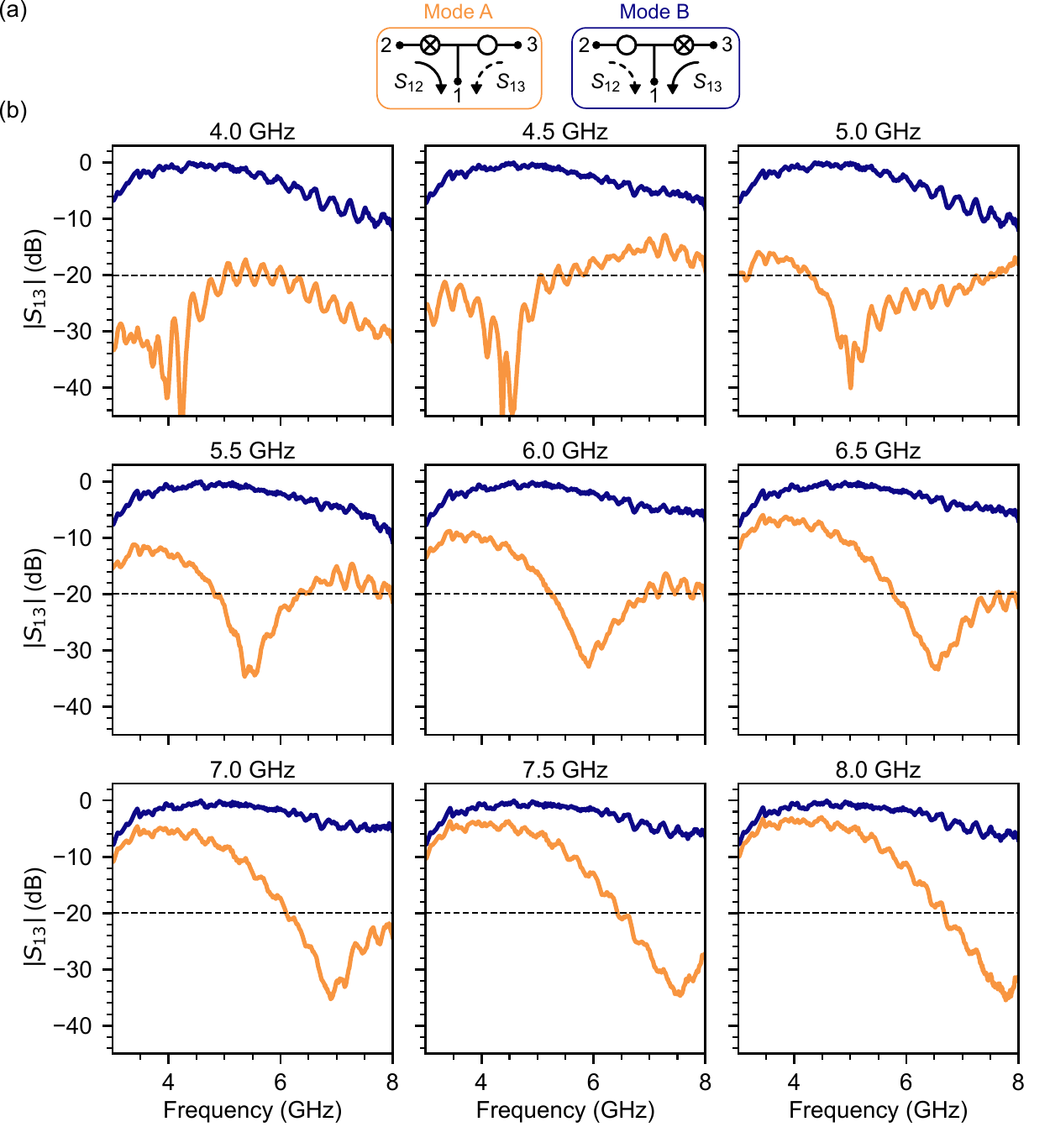}
	\end{center}
	\caption{
		\label{fig:s13all}
		Coupler performance for different operating frequencies, between ports 1 and 3.
		(a) Schematic of the two modes measured here; see main text.
		(b) Measured transmission from port 3 to port 1 $(|S_{13}|)$ as a function of frequency for mode A (orange) and mode B (blue), demonstrating broadband operation. Dashed line represents the $-20$ dB transmission threshold used to define the isolation bandwidth.
	}
\end{figure}

\renewcommand{\baselinestretch}{1.0}
\begin{table}
	\begin{center}
		\begin{tabular}{|c|c|c|c|}
			\hline
			\hline
			Operating frequency (GHz) & On/off ratio (dB) & Transmission bandwidth (GHz) &  Isolation bandwidth (GHz) \\
			$f_0$ & $|S_{12}^{A}/S_{12}^{B}|$ & $|S_{12}^{A}| > -1$ dB & $|S_{12}^{B}| < -20$ dB \\
			\hline
			4.025 & 43.423 & 1.28 & $>$ 1.995 \\
			4.405 & 61.48 & 1.595 & 2.465 \\
			5.085 & 46.899 & 1.48 & $>$ 3.8 \\
			5.565 & 34.574 & 1.6 & 2.39 \\
			6.135 &  34.19 & 1.595 & 2.62 \\
			6.455 &  33.72 & 1.24 & 2.18 \\
			7.03 &  33.614 & 1.485 & $>$ 1.9 \\
			7.68 &  33.845  & 1.59  & $>$ 1.58 \\
			7.965 & 35.025 & 1.595 & $>$ 1.3 \\
			\hline
			\hline
			Operating frequency (GHz) & On/off ratio (dB) & Transmission bandwidth (GHz) & Isolation bandwidth (GHz) \\
			$f_0$ & $|S_{13}^{B}/S_{13}^{A}|$ & $|S_{13}^{B}| > -1$ dB & $|S_{13}^{A}| < -20$ dB \\
			\hline
			4.225 & 49.529 & 1.58 & $>$ 1.995 \\
			4.37 & 49.134 &  1.56 & $>$ 2.315 \\
			5.005 & 40.066 & 1.58 & 3.22 \\
			5.36 & 34.66 & 1.545 & 1.51 \\
			5.915 & 32.902 & 1.54 & 1.685 \\
			6.57 & 33.393 & 1.55 & 1.845 \\
			6.895 & 35.285 & 1.59 & $>$ 1.9 \\
			7.54 & 34.686 & 1.535 & $>$ 1.57 \\
			7.775 & 35.487 & 1.455 & $>$ 1.32 \\
			\hline			
			\hline
		\end{tabular}
	\end{center}
	\caption{\label{table:performance}
		Performance of the variable coupler at different operating frequencies, measured for $S_{12}$ (upper half) and $S_{13}$ (lower half). All parameters are calculated from the spectroscopy data shown in Fig.~\ref{fig:s12all} and Fig.~\ref{fig:s13all}.}
\end{table}
\renewcommand{\baselinestretch}{1.5}
